# On the Probability of "The Inhomogeneity of Composition Along the Magnetic Cloud Axis"

Yu. I. Yermolaev [a], * A.A. Khokhlachev [a] and I. G. Lodkina [a]

[a] *Space Research Institute, Russian Academy of Sciences, Moscow, Russia*
*e-mail: yermol@iki.rssi.ru*

## Abstract

In the present work, we study the possibility of assessing the inhomogeneities of the ionic composition along the axis of the magnetic cloud using the method that was recently used by Song et al. (2021). Possible violations of the used assumptions do not allow one to draw reliable conclusions.

## Article

Studies of properties of coronal mass ejection (CME) in interplanetary space are one of the most important problems in CME physics and space weather (e.g., Gopalswamy, 2016; Temmer, 2021; Yermolaev et al., 2021 and references therein). In recent paper, Song et al. (2021) studied an magnetic cloud (MC) that was observed by Advanced Composition Explorer at ~ 1 AU during March 4–6, 1998, and Ulysses at ~ 5.4 AU during March 24–28, 1998, sequentially, and concluded that the MC has "The Inhomogeneity of Composition Along the Magnetic Cloud Axis". However, the authors did not take into account the fact that a number of the assumptions made may not be fulfilled, and if they are violated, the conclusions drawn turn out to be unreliable. Here we discuss the assumptions that, in our opinion, can be violated.

Further in our article, we will use the more general term Interplanetary Coronal Mass Ejection (ICME) for interplanetary manifestations of CMEs, magnetic cloud is a particular type of ICMEs. Experimental evidence suggests that distinctions between MC and ejecta (non-MC) can be partially connected to the conditions of observations (distance between axis of ICME and spacecraft trajectory), but not to physical distinctions between ejecta and MCs (e.g., Jian et al., 2006; Yermolaev et al., 2017 and references therein).

First of all, it should be noted that the authors implicitly assume that the distribution of parameters across the MC cross section is uniform, and the difference in spacecraft trajectories of several degrees relative to the MC axis will not lead to differences in the measured values. However, when measuring the similar non-uniform distributions of parameters, real differences in the spacecraft trajectories lead to differences in measurements and to incorrect conclusions. As the results of statistical analysis of ~ 1600 events near the Earth in the period 1976-2017 showed, ICMEs contain an electric current with an increased helium abundance, and this current with a cross-sectional dimension of ~ $10^6$ km has about 10% of the ICMEs size, and the abundance of helium outside the electric current is close to one outside ICMEs (Yermolaev, 2019; Yermolaev et al., 2020). At least, the abundance of helium has an uneven distribution in the cross section of the MC. It is natural to assume that the distributions of other parameters are also uneven in this plane. According to data in the paper "both spacecraft were located around the ecliptic plane, and the latitudinal and longitudinal separations between them were ~2.2° and ~5.5°, respectively". In our opinion, with such differences in trajectories and at unknown distances of the trajectories from the MC axis, it is impossible to make a reliable conclusion based on 1 event that the inhomogeneity is distributed along the MC axis. It is more likely that inhomogeneity is observed across the axis of the MC.

It should also be noted that the statement made in the paper that "The composition (including the ionic charge states and elemental abundances) is determined prior to and/or during CME eruptions in the solar atmosphere and does not alter during MC propagation to 1 AU and beyond" is true only if the plasma volume does not contain parameter gradients on the scales under study. Otherwise, diffusion and plasma instabilities tend to reduce this gradient and equalize the parameter values throughout the volume (e.g., Zelenyi & Milovanov, 2004; Yermolaev, 2014 and references therein). Since the helium abundance inside ICMEs has a pronounced gradient, its distribution in the MC volume can significantly change over time with distance from the Sun. For example, at the speed of alpha-particles with an Alfvén speed relative to the main, proton, component of the solar wind (e.g., Asbridge et al., 1976 ; Bosqued et al., 1977; Neugebauer, 1981; Marsch et al., 1982; Ogilvie et al., 1982) during the MC movement from 1 to 5 AU the helium component can leave the volume of the electric current or even the volume of the entire MC.

There is experimental evidence that an increase in the abundance of helium is recorded in plasma discontinuities of the solar wind (at small scales of ~ $10^4$ km or less) (Borovsky, 2008; Šafránková et al., 2013; Sapunova et al., 2020), and it is suggested that there are some universal mechanisms that involve heavy ions in current sheets (Grigorenko et al., 2017; Yermolaev et al., 2020). However, it is not known whether such mechanisms exist in ICMEs on scales that are 2-3 orders of magnitude larger than the scales in solar wind discontinuities.

Thus, the approach used by Song et al. (2021) for the analysis of a single event, contains incorrect assumptions and does not allow obtaining data that could unequivocally testify in favor of the statement "The Inhomogeneity of Composition Along the Magnetic Cloud Axis", and the conclusions drawn in the work are not supported by experiment and have an extremely low probability really be realized.